\documentclass[twocolumn,aps,prd,nofootinbib,amsmart]{revtex4-2}
 \usepackage{graphicx,amssymb}
 \usepackage[utf8]{inputenc} % Required for inputting international characters

\usepackage{comment}
\usepackage[urlcolor=blue,colorlinks,breaklinks=true]{hyperref}

\PassOptionsToPackage{obeyspaces,spaces,hyphens}{url}

 \usepackage{amsmath}
 \usepackage{svg}
 \begin{document}
 	%My commands
 	\def\half{{1\over2}}
 	\def\shalf{\textstyle{{1\over2}}}
 	
 	\newcommand\lsim{\mathrel{\rlap{\lower4pt\hbox{\hskip1pt$\sim$}}
 			\raise1pt\hbox{$<$}}}
 	\newcommand\gsim{\mathrel{\rlap{\lower4pt\hbox{\hskip1pt$\sim$}}
 			\raise1pt\hbox{$>$}}}

\newcommand{\be}{\begin{equation}}
\newcommand{\ee}{\end{equation}}
\newcommand{\bq}{\begin{eqnarray}}
\newcommand{\eq}{\end{eqnarray}}
 	
\title{Domain walls in light of Cosmic Microwave Background and Pulsar Timing Array data}

\author{D. Gr{\"u}ber}
\email[Electronic address: ]{david.grueber@astro.up.pt}
\affiliation{Departamento de F\'{\i}sica e Astronomia, Faculdade de Ci\^encias, Universidade do Porto, Rua do Campo Alegre 687, PT4169-007 Porto, Portugal}
\affiliation{Instituto de Astrof\'{\i}sica e Ci\^encias do Espa{\c c}o, Universidade do Porto, CAUP, Rua das Estrelas, PT4150-762 Porto, Portugal}

\author{P.P. Avelino}
\email[Electronic address: ]{pedro.avelino@astro.up.pt}
\affiliation{Departamento de F\'{\i}sica e Astronomia, Faculdade de Ci\^encias, Universidade do Porto, Rua do Campo Alegre 687, PT4169-007 Porto, Portugal}
\affiliation{Instituto de Astrof\'{\i}sica e Ci\^encias do Espa{\c c}o, Universidade do Porto, CAUP, Rua das Estrelas, PT4150-762 Porto, Portugal}
\affiliation{Université Côte d’Azur, Observatoire de la Côte d’Azur, CNRS, Laboratoire Lagrange, France}

\author{L. Sousa}
\email[Electronic address: ]{lara.sousa@astro.up.pt}
\affiliation{Instituto de Astrof\'{\i}sica e Ci\^encias do Espa{\c c}o, Universidade do Porto, CAUP, Rua das Estrelas, PT4150-762 Porto, Portugal}
\affiliation{Departamento de F\'{\i}sica e Astronomia, Faculdade de Ci\^encias, Universidade do Porto, Rua do Campo Alegre 687, PT4169-007 Porto, Portugal}

\date{\today}

\date{\today}
\begin{abstract}

In this paper, we study the compatibility of biased domain wall scenarios with current gravitational wave data. We show that the Cosmic Microwave Background bounds on the fractional density of gravitational waves at the time of decoupling may only slightly improve on the constraints that result from requiring that domain walls never dominate the cosmic energy budget. We show that, despite this, the range of energy scales of the domain-wall forming phase transitions are already quite constricted, even if the networks decay early in cosmological history. We also show that, if domain walls are to provide an explanation to the stochastic gravitational wave background that was recently detected by pulsar timing arrays, they not only have to decay early in the radiation dominated era but also their energy density would have to be close to dominating the energy density of the universe, which would require some fine tuning of the parameters of the models.

\end{abstract}

\maketitle

\section{Introduction}
The recent confirmation of the detection of a stochastic gravitational wave background (SGWB) by Pulsar Timing Arrays (PTAs)~\cite{NANOGrav:2023gor,Reardon:2023gzh,EPTA:2023fyk,Xu:2023wog} has ignited speculations about its origin. While astrophysical phenomena are considered a plausible explanation for the detected signals~\cite{NANOGrav:2023gor}, the possibility of a cosmological source remains open. Among those, Gravitational Waves (GWs) from inflation, phase transitions in the early universe, or topological defect networks, such as cosmic strings, are frequently explored in the literature~\cite{NANOGrav:2023hvm}. %In this paper we will discuss the possibility of the PTA signals being sourced by cosmic domain walls.

Like other cosmic defects, domain walls may arise as a consequence of the spontaneous breaking of a symmetry associated with a scalar field~\cite{vilenkinshellardCosmicStringsOther}. Domain walls, however, if they are to survive until the present day, are tightly constrained by the Zel'dovich bound, which essentially limits the energy scale of the wall-forming phase transition to be below approximately 1 MeV~\cite{Zeldovich:1974uw}. Domain wall networks, however, may not be inherently stable. Certain models, such as those linked to axion-like particles~\cite{Huang:2024nbd,Ferreira:2024eru,Dunsky:2024zdo,Kitajima:2023cek,Blasi:2022ayo} --- which are receiving increasing interest in the literature --- suggest that domain wall networks may have decayed in the past due to a bias in their scalar field potential. In this case, domain wall networks would evade this stringent bound and could be formed much earlier in cosmological history. Currently, constraints on stable domain wall network scenarios are mainly based on the Cosmic Microwave Background (CMB)~\cite{Sousa:2015cqa}, but they should also emit gravitational radiation throughout their evolution. Their GW signatures should then allow us to constrain particle physics scenarios.\\A framework was recently developed in~\cite{Gruber:2024dgw}, where the authors developed a semi-analytical model to characterize the SGWB produced by domain walls. In the present work, we will build upon this work to constraint the energy scale at domain wall formation as a function of the time of their decay, by confronting this model with observational constraints on the SGWB at the time of decoupling from the CMB. We also explore the possibility of domain walls providing a major contribution to the SGWB that was recently detected by PTAs and derive the region of parameter space that is consistent with these observations.\\In section \ref{sec:SGWBWalls} we will briefly recap the framework presented in ~\cite{Gruber:2024dgw} that describes the SGWB by domain walls. In section \ref{sec:CMBConstraints} we will then show how the constraint on the number of relativistic particle species at the time of decoupling can be used to constraint the SGWB from domain wall networks and introduce some further, more general constraints. We will apply the aforementioned framework in section \ref{sec:PTAdetection} to the recent PTA signal to investigate the possibility of the latter stemming from biased domain walls. Then we conclude in section \ref{sec:Conclusion}.\\Throughout this paper, we will work in natural units, where $c=\hbar=1$, with $c$ being the speed of light in vacuum and $\hbar$ being the reduced Planck constant. In these units Newton's gravitational constant is given by $G=6.70711\cdot 10^{-57} \, \text{eV}^{-2}$. Moreover, we will use the cosmological parameters measured by the Planck mission \cite{planckcollaborationPlanck2018Results2020}, where the values of the density parameters for radiation, matter and dark energy at the present time are respectively given by $\Omega_{\rm r} = 9.1476\cdot 10^{-5}$, $\Omega_{\rm m} = 0.308$, $\Omega_\Lambda=1-\Omega_{\rm r} - \Omega_{\rm m}$ and the Hubble parameter is $H_0=2.13 \cdot h \cdot 10^{-33} \, \rm eV$, with $h=0.678$.

\section{SGWB generated by domain wall networks: an overview}
\label{sec:SGWBWalls}
In this section, we provide a brief overview of the stochastic gravitational wave background generated by domain walls, which was studied in detail in~\cite{Gruber:2024dgw}. This spectrum is sourced by the continuous collapse of domain walls that detach from the network as a result of the expansion of the background. However, since the gravitational wave emission in the late stages of domain wall collapse is not yet well understood (as numerical simulations are unable to resolve the domain walls in the ultrarelativistic limit), some details of the power spectrum are uncertain. In fact, in~\cite{Gruber:2024dgw}, the SGWB spectrum depends on two unknown parameters. The total energy density of GWs emitted, however, may be computed with minimal assumptions and, for this reason, it will serve as the basis for our analysis in this paper.

\subsection{GW energy density}
\label{sec:GWsEnDens}
In \cite{Gruber:2024dgw} the fractional contribution of the SGWB produced by standard and biased domain wall networks to the cosmic energy budget has been estimated, assuming that a constant fraction $\mathcal F$ of the energy of collapsing domain walls is emitted in the form of GWs. This has been achieved by resorting to the Velocity-dependent One-Scale (VOS) model~\cite{avelinoDomainWallNetwork2011,Sousa:2011ew,Sousa:2011iu} to describe the cosmological evolution of their characteristic scale $L \equiv \sigma/\rho=\eta^3/\rho$ --- where $\sigma$, $\eta \equiv \sigma^{1/3}$ and $\rho$ are, respectively, the domain wall tension, their characteristic energy scale and the average energy density of the domain wall network ---  and root-mean-squared velocity $\bar v$. 

For a power-law evolution of the scale factor --- so that $a \propto t^\lambda$ with $\lambda=1/2$ and $\lambda=2/3$ in the radiation and matter eras, respectively --- the (biased) domain wall network evolves in a linear scaling regime of the form $L=\xi t$ and $\bar v = \rm const$ until its eventual decay at a time $t_\star$ (corresponding to a scale factor $a_\star$ and a redshift $z_\star=1/a_\star-1$). In this case, one should have that~\cite{Gruber:2024dgw}
\be 
\Omega_{\rm gw}\equiv \left.\frac{8 \pi G \rho_{\rm gw}}{3 H^2}\right|_{t=t_0}=\sigma \mathcal{F} C_1 (1+z_\star)^{-2} \mathcal{G}_\star^{1/2}    \,,\label{OGWwalls}
\ee
for $z_\star \le z_{\rm eq}$, and 
\bq
\Omega_{\rm gw} &=& \sigma \mathcal{F}\mathcal{G}_\star^{1/2} \nonumber\\
&\times& \left[C_2 \left( (1+z_\star)^{-\frac{5}{2}} - (1+z_{\rm eq})^{-\frac{5}{2}} \right) + C_3\right] \,,
\label{eq:constraintMatter}
\eq
for $z_\star>z_{\rm eq}$. Here, $z_{\rm eq}=\Omega_m/\Omega_r-1$ is the redshift of radiation-matter equality, 
\bq
C_1 &=& \frac{16\pi G \Omega_r^{\frac{1}{2}}\mathcal{A}_r}{3 H_0 }\,, \\ 
C_2 &=& \frac{12 \pi G \Omega_m^\frac{1}{2} \mathcal{A}_m}{5 H_0} \,, \\
C_3&=&C_1\left(\frac{\Omega_r}{\Omega_m}\right)^2\,,
\eq
and $\mathcal{A}_{\rm i} \equiv \tilde{c}\bar{v}_{\rm i} / \xi^2_{\rm i}$. The values of the dimensionless  linear scaling parameters $\xi$ and $\bar v$ are respectively given by $\xi_{\rm r} =1.532$, ${\bar v}_{\rm r}=0.479$, and $\xi_{\rm m} =1.625$, ${\bar v}_{\rm m}=0.339$ in the radiation and matter eras for $\tilde c = 0.5$ (an excellent approximation in both eras~\cite{leiteScalingPropertiesDomain2011}). The factor of $\mathcal{G}(z)$ is a small correction to the result derived in~\cite{Gruber:2024dgw} that accounts for the impact of the change in the effective number of massless degrees of freedom on the expansion rate. Here, we follow the approach in Ref.~\cite{Binetruy:2012ze} and approximate the Hubble parameter by 
\be
H(z)=H_0\left[\Omega_\Lambda+\Omega_{\rm m}(1+z)^3 +\mathcal{G}(z) \Omega_{\rm r}(1+z)^4 \right]^{1/2}\,,
\ee
where
\be
\mathcal{G}(z) = \begin{cases}
1  & \text{for }  z < z_{\rm e^-e^+} \\
0.83  & \text{for }  z_{\rm e^-e^+} < z <  z_{\rm QCD} \\
0.39  & \text{for }  z > z_{\rm QCD} 
\end{cases}
\ee
describes the decreases in the number of degrees of freedom expected to occur, in the Standard Model of Particle Physics, at the QCD phase transition and as a result of the electron-positron annihilation (throughout this paper we shall take $z_{\rm e^-e^+} = 10^{9}$ and $z_{\rm QCD} = 10^{12}$ \cite{Binetruy:2012ze}). This factor is responsible for a  suppression of the fractional contribution of the gravitational waves emitted by domain wall networks to the cosmic energy budget by a factor of $\mathcal{G}_\star^{1/2}\equiv \mathcal{G}(z_\star)^{1/2}$.

Notice that, throughout this paper we shall assume that the network is in a linear scaling regime before it is rendered unstable and instantaneously decays without producing further GWs. Including the potential contribution of the decay of the network to the SGWB could increase its amplitude by up to an order of magnitude~\cite{Gruber:2024dgw} (but there is still some uncertainty regarding this contribution). We will briefly comment on the impact of this assumption in the discussion of this paper.

\subsection{Power spectrum}
\label{sec:GWsSpec}
The power spectrum of the SGWB generated by domain walls, characterized by
\be 
\Omega_{\rm gw}(f)= \frac{8\pi G}{3H_0^2}\frac{d\rho_{\rm gw}}{d\log f}\,,
\ee
can be roughly approximated by a broken power-law shape~\cite{Gruber:2024dgw,hiramatsuEstimationGravitationalWave2014}, with the power being a monotonically increasing (decreasing) function of the frequency for frequencies smaller (larger) than the peak frequency. The frequency of the peak of the spectrum is expected to be inversely proportional to the characteristic length of the network at the time of decay of the network, and is thus roughly given by\footnote{Aside from $\mathcal{O}(1)$ factors associated to the changes in the number of relativistic degrees of freedom and to the uncertainties regarding the power spectrum~\cite{Gruber:2024dgw}.}
\be
f_{\rm p} = \frac{2}{\alpha}  \frac{a_\star} {L_\star}\,,
\label{eq:fpeak}
\ee
where $L_\star=\xi_\star t_\star=\xi \lambda_\star H_\star^{-1}$ and $\alpha<1$ is a parameter introduced to characterize the typical scale of GW in terms of $L_\star$ (see \cite{Gruber:2024dgw} for more details). Most of the GW energy density is, in general, concentrated within a narrow frequency range around the peak of the spectrum and, in fact, the peak amplitude is, aside from a factor that is generally of order unity, approximately given by Eqs.~(\ref{OGWwalls})-(\ref{eq:constraintMatter}).

\section{Cosmic Microwave Background constraints}
\label{sec:CMBConstraints}
The domain wall network is expected to have been formed at a temperature $T_{\rm birth} \sim \eta$. Since the network (obviously) cannot decay before it is formed, the characteristic energy scale of a domain wall network that survives until $z_\star$ necessarily has to fulfill the condition $T_\star < T_{\rm birth}$. We should then have that:
\be
\eta > \left(\frac{45}{4\pi^3} \frac{H_0^2 \mathcal{G}_\star \Omega_{\rm r}}{G g_\star} \right)^{1/4}(1+z_\star) \,, \label{birth}
\ee
where $g_\star=g(z_\star)$ and~\cite{Binetruy:2012ze}
\be
g(z) = \begin{cases}
3.36  & \text{for }  z < z_{\rm e^+e^-} \\
10.75  & \text{for }  z_{\rm QCD} < z < z_{\rm e^+e^-} \\
106.75  & \text{for }  z > z_{\rm QCD}
\end{cases}\,,
\ee
describes the effective number of degrees of freedom at a redshift $z$.

Moreover, domain walls should always provide a subdominant contribution to the cosmic energy budget. This condition results in an additional constraint on their characteristic energy scale of the form:
\begin{equation}
     \eta < \left( \frac{3H_\star \lambda_\star \xi_\star}{8\pi G}\right)^{1/3}\, .
    \label{eq:rhoSmallerroCrit}
\end{equation}
Here, $\lambda_\star=\lambda_{\rm r}$ and $\xi_\star=\xi_{\rm r}$ (or $\lambda_\star=\lambda_{\rm m}$ and $\xi_\star=\xi_{\rm m}$) for $z_\star$ greater (or smaller) than $z_{\rm eq}$.

Finally, the power spectrum of temperature and polarization anisotropies of the CMB is extremely sensitive to the expansion rate at the time of decoupling, which, for a flat universe, is fully determined by the energy density at that time. As a result, CMB observations impose stringent constraints on the effective number of relativistic degrees of freedom at the time of decoupling, which provides a tight limit on the total energy density in gravitational waves generated before that time. One recently derived constraint that takes into account the Planck 2018 data~\cite{planckcollaborationPlanck2018Results2020} places this limit at 

\be 
\Omega_{\rm gw}^{\rm (CMB)} h^2 < 1.7\cdot 10^{-6}
\label{eq:CMBconstraint}
\ee
for frequencies $f\gtrsim 10^{-15}\, \rm Hz$ \cite{Clarke:2020bil}. There are also similar constraints at the time of the primordial nucleosynthesis, but these apply only at much larger redshifts ($z \gsim 10^9$).

\begin{figure} 
	         \begin{minipage}{1.\linewidth}  
               \rotatebox{0}{\includegraphics[width=1\linewidth]{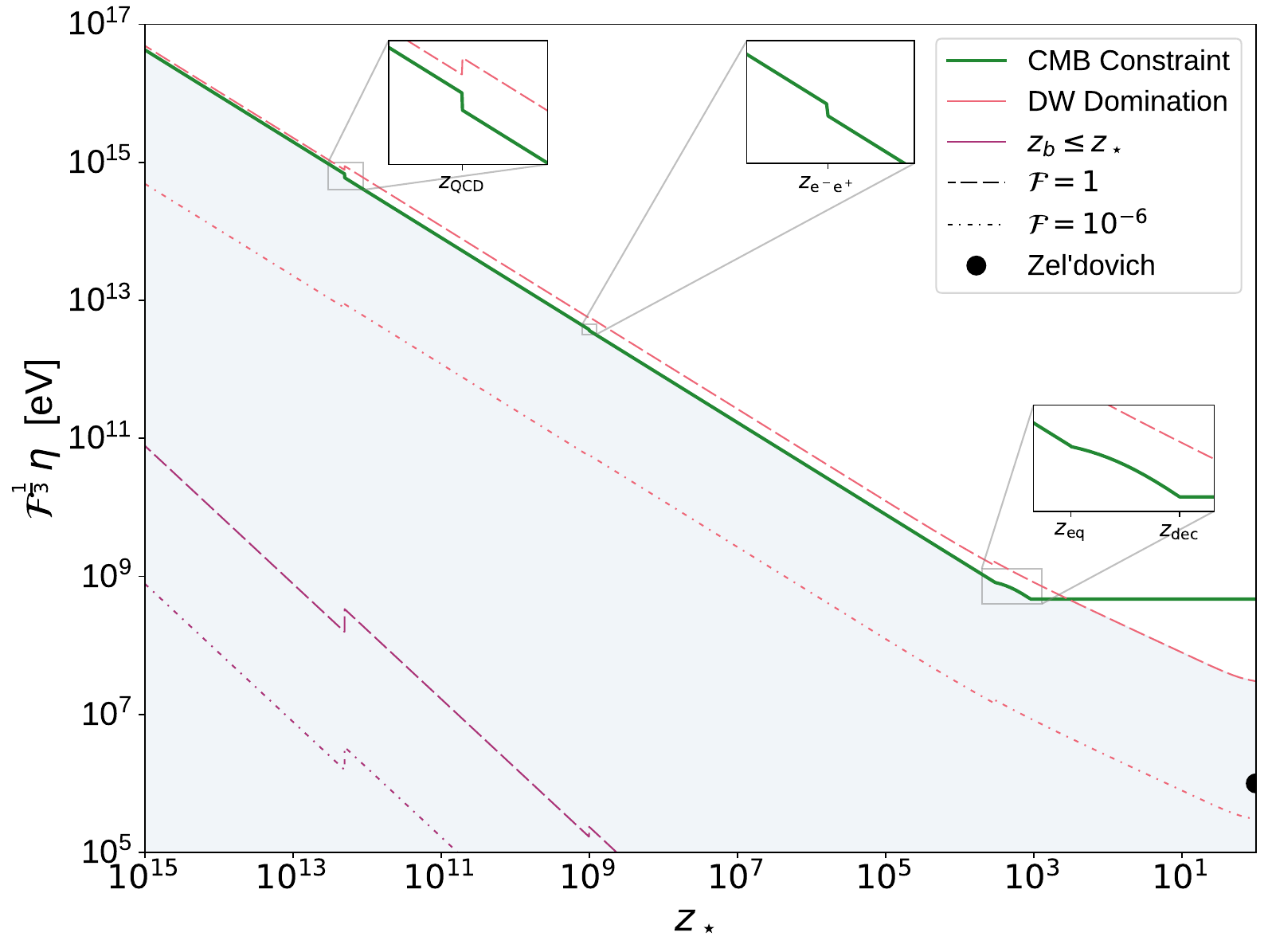}}
               %\rotatebox{0}{\includesvg[width=1\linewidth]{./figures/Constraint_CMB.svg}}
         \end{minipage}
	\caption{Combined constraints on $\mathcal{F}^{1/3}\eta$ as a function of the redshift $z_\star$ at the time of decay of the domain wall network. The green line represents the upper limit on $\mathcal{F}^{1/3}\eta$ resulting from the CMB bound in Eq.~(\ref{eq:CMBconstraint}). The red lines represent the limit in which domain walls dominate the universe's energy budget (area above is excluded), while the purple lines mark the boundary where the time of decay coincides with the time of formation of the domain wall network (area below is excluded). In the two latter cases, the dashed- and dashed dotted lines correspond to values of $\mathcal{F}=1$ and $\mathcal{F}=10^{-6}$ respectively. The Zel'dovich bound is indicated by the black marker in the plot. The shaded area represents the allowed region in parameter space if $\mathcal{F}$ is treated as a free parameter.} 
	\label{fig:constraintsCMB}
\end{figure}

In Fig.~\ref{fig:constraintsCMB}, we display the values of $\mathcal{F}^{1/3} \eta$ allowed by CMB data (shaded area) as a function of the redshift of decay of the network. This allowed region of the $(z_\star,\mathcal{F}^{1/3} \eta)$-parameter space is obtained by combining the bound obtained by requiring that $\Omega_{\rm gw}(z_{\rm dec})\le \Omega_{\rm gw}^{\rm (CMB)}$ (green line) --- where $z_{\rm dec}$ is the redshift at the time of decoupling --- and that domain walls never dominate the energy budget (red lines). Regarding the latter bound, we include the cases in which $\mathcal F =1$ (dashed line) and $\mathcal F =10^{-6}$ (dot-dashed line). For $\mathcal{F}=1$, this actually corresponds to an absolute constraint and it is only slightly less stringent than the upper bound on $\mathcal F^{1/3} \eta$ obtained from the CMB bound on $\Omega_{\rm gw}$ for $z<z_{\rm dec}$. However, since the CMB constrains only the total energy density of GWs created until decoupling, the resulting bounds for walls that survive past decoupling are independent of $z_\star$ and weaker than those that result from imposing that domain walls are always subdominant for $\mathcal{F}=1$. For values of $\mathcal{F}\lesssim 0.3$, Eq. \eqref{eq:rhoSmallerroCrit} is always more constraining than the CMB bound. 

One may also derive lower bounds on $\mathcal{F}^{1/3}\eta$, using Eq.~(\ref{birth}), but these also depend on $\mathcal{F}$. These bound are included in Fig.~\ref{fig:constraintsCMB}, as dashed and dot-dashed purple lines for $\mathcal F=1$ and $\mathcal{F}=10^{-6}$, respectively. Notice that, for a given value of $\mathcal F$, the allowed region of parameter space is also bounded from below by this constraint and then is given by a narrow band.

\section{Domain wall interpretation of PTA data}\label{sec:PTAdetection}

Recently, different PTAs --- including the North American Nano-hertz Observatory for Gravitational waves (NANOGrav)~\cite{NANOGrav:2023gor}, the Parks PTA (PPTA)~\cite{Reardon:2023gzh}, the European PTA (EPTA)~\cite{EPTA:2023fyk} and the Chinese PTA (CPTA)~\cite{Xu:2023wog} --- have produced the first convincing evidence for a SGWB in the nanohertz frequency band. These results all indicate the existence of an excess signal with a characteristic strain amplitude of the order $\mathcal{O}(10^{-15})$ at a reference frequency of $1 \, {\rm yr}^{-1}\simeq 31.7 \, {\rm nHz}$. Assuming that the spectral index of the spectrum $d$ --- defined such that $\Omega_{\rm gw} \propto f^d$ locally --- is free, this data may be translated into the following value for the spectral density of gravitational waves:
\be 
\Omega_{\rm gw}(f)h^2\simeq 6.3\times 10^{-10} A^2 \left(\frac{f}{{\rm yr}^{-1}}\right)^d \,,
\label{eq:omegaPTA}
\ee 
with a range of allowed parameters of $A=6.4^{+4.2}_{-2.7}$ and $d=1.8\pm 0.6$ for NANOGrav~\cite{NANOGrav:2023gor}, $A=2.9^{+2.6}_{-1.8}$ and $d=0.81^{+0.63}_{-0.73}$ for EPTA~\cite{EPTA:2023fyk} and $A=3.1^{+1.3}_{-0.9}$ and $d=1.1\pm 0.4$ for PPTA~\cite{Reardon:2023gzh}. This spectral shape is valid within a frequency range of roughly $[f_{\rm min},f_{\rm max}]$, with $f_{\rm min} = 3.2\cdot 10^{-8} \, \rm Hz$ and $f_{\rm max} = 10^{-7} \, \rm Hz$. Here, we will investigate whether such a signature could be produced by a biased domain wall network as was proposed in the literature~\cite{Ferreira:2024eru,NANOGrav:2023hvm,Kitajima:2023cek,Zhang:2023nrs,Gouttenoire:2023ftk}. 

Given the uncertainty associated to the shape of the SGWB power spectrum, here we will again focus on the total gravitational wave energy density instead. A conservative lower bound on the fractional contribution of the SGWB to the cosmic energy budget in the PTA frequency range may be obtained by calculating
\be
\Omega_{\rm gw}^{\rm (PTA)}=\int_{f_{\rm min}}^{f_{\rm max}} \Omega_{\rm gw}(f) \frac{df}{f}\,,
\ee
while considering the lower bounds on the parameters $A$ and $d$. This gives us the minimum GW energy density one would need to have in this range to generate the observed signal in each PTA experiment: $\left( \Omega_{\rm gw}^{\rm (NANO)},\, \Omega_{\rm gw}^{\rm (EPTA)},\, \Omega_{\rm gw}^{\rm (PPTA) } \right) = \left( 4.643\cdot 10^{-8},\, 1.995\cdot 10^{-9},\,1.170\cdot 10^{-8}\right)$.

The observed PTA spectrum, fitted by the function given in Eq. \eqref{eq:omegaPTA}, appears to be an increasing function of frequency, as all experiments predict that $d>0$. Hence, if the main contribution to this spectrum was to be provided by the gravitational radiation produced by biased domain walls, it should mostly correspond to the part of the spectrum that lies to the left of the peak (on the lower frequency side). Moreover, since for $f\ll f_{\rm p}$, one should have $\Omega_{\rm gw}(f) \propto f^3$ due to causality constraints~\cite{Caprini:2009fx}, the peak of the spectrum cannot be located at a frequency that is much larger than $f_{\rm max}$. On one hand, PTA data seems to be consistent with smaller values of $d$ --- which would indicate that, if domain walls are to be the main source, the detected signal should correspond to the transitional region between the peak and the regime characterized by $d=3$. On the other hand, if $f_{\rm p}\gg f_{\rm max}$, the amplitude of the peak would have to be very large in order to explain the signal within the PTA window and $\Omega_{\rm gw}$ would then exceed the stringent bounds discussed in the previous section. In fact, if the spectrum between $f_{\rm max}$ and $f_{\rm p}$ has a spectral index of $d=3$, the total energy contained in the frequency interval $[f_{\rm max},\,f_{\rm p}]$ would exceed the total energy limits dictated by the CMB bound for $f_{\rm p}\gsim 5\cdot10^{-7} \, \rm Hz = 5\cdot f_{\rm max}$. The upper limit we compute for the redshift $z_\star$ at the time of decay of the network, under the assumption that $f_{\rm p}$ is contained in $[f_{\rm min},f_{\rm max}]$, may then increase at most by this factor of 5. We will then assume that the peak of the spectrum lies within the frequency interval $[f_{\rm min},f_{\rm max}]$.

\begin{figure} 
	         \begin{minipage}{1.\linewidth}  
               \rotatebox{0}{\includegraphics[width=1\linewidth]{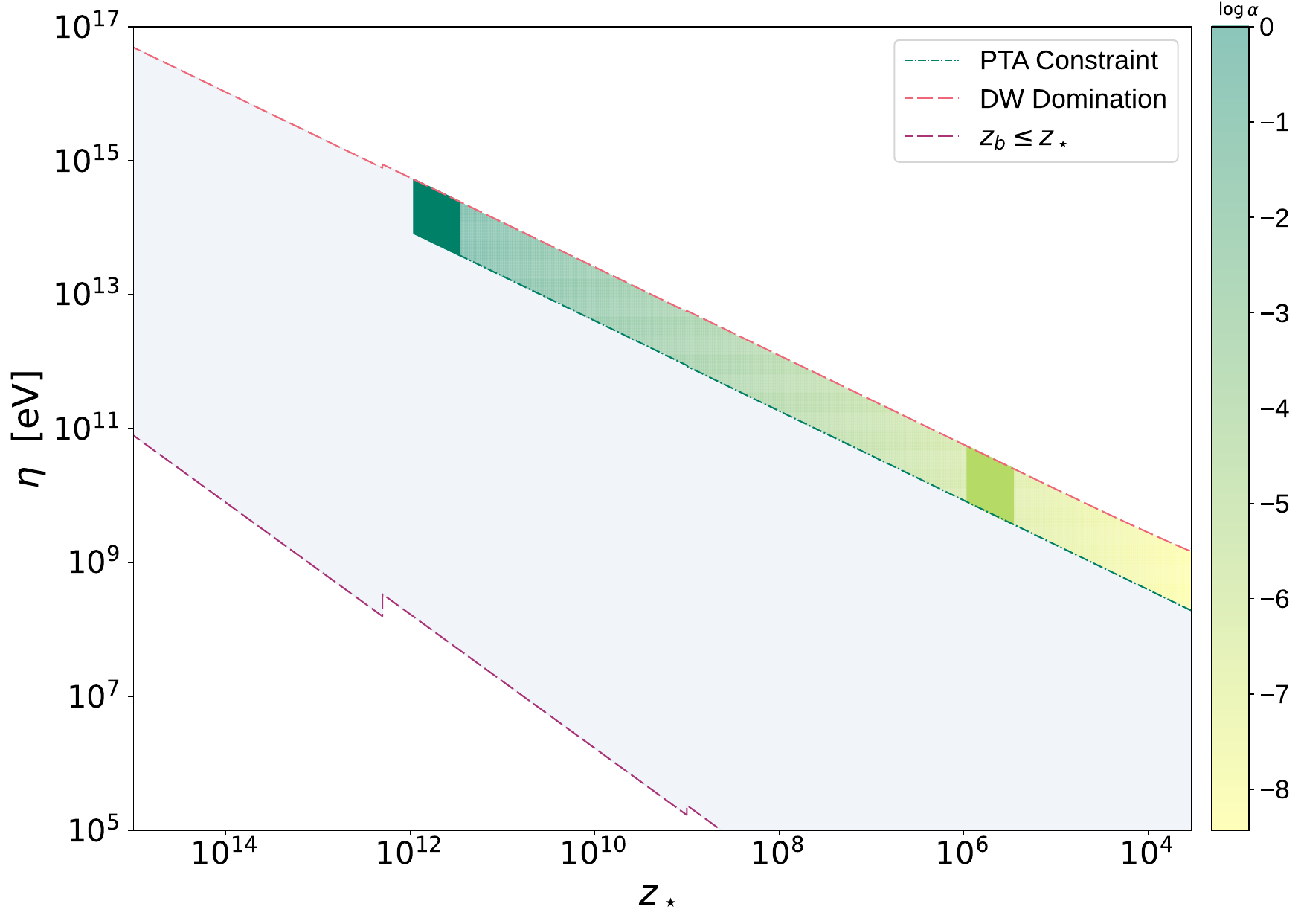}}
         \end{minipage}
	\caption{Allowed values for the characteristic energy scale $\eta$ of domain walls as a function of the redshift of decay $z_\star$. The red and purple lines represent the constraints resulting from imposing that walls cannot dominate the energy budget of the universe and that their time of decay cannot be before the time of their formation. The area shaded in blue represents the region of parameter space compatible with these constraints. The color shaded band represents the region of parameter space compatible with the NANOGrav data, assuming domain wall networks constitute the main contribution to this signal. The dark- and light green rectangles represent the allowed regions assuming a fixed value of $\alpha=1$ and $\alpha=10^{-6}$ respectively. More details in the text.}
	\label{fig:constraintsPTA}
\end{figure}

If $f_{\rm p}  \gsim f_{\rm min}$ and assuming that $\alpha \gsim 10^{-8}$, then the domain wall network must have decayed during the radiation dominated era. Eq.~(\ref{eq:fpeak}) then implies
\be
f_{\rm p} = \frac{8 H_0 \Omega_r^\frac{1}{2}}{\alpha \xi_r}(1+z_\star) \,. \label{fp}
\ee
The redshift of decay of the network may then be inferred from the location of the peak for each value of $\alpha$. Then, using Eq.~(\ref{OGWwalls}) and the lower bounds we have derived for the value of $\Omega_{\rm gw}$ in different PTA experiments, we may derive constraints on $\mathcal{F}^{1/3} \eta$. Note that, in deriving these constraints, we will neglect the contribution to $\Omega_{\rm gw}$ outside of the $[f_{\rm min},f_{\rm max}]$ range and therefore the bounds derived here should be conservative. Moreover, since the domain wall SGWB should be simultaneously consistent with the data of all PTA experiments, we consider $\Omega_{\rm gw}^{\rm (NANO)}=4.64\cdot 10^{-8}$, which corresponds to the strongest signal detected across the various PTA experiments.

In Fig. \ref{fig:constraintsPTA}, we plot the allowed region of the $(z_\star,\eta)$-parameter space (shaded blue area) for domain wall networks. This figure includes the constraints in Eqs.~\eqref{birth} and~\eqref{eq:rhoSmallerroCrit} --- again plotted as red and purple lines respectively --- but here they become fixed boundaries of the allowed parameter space, since we display this only in terms of the energy scale $\eta$ (i.e., independently of $\mathcal{F}$).  Additionally, the green dashed line marks the minimal characteristic energy scale $\eta$ required for domain wall networks to generate the PTA signals for $\mathcal{F}=1$. Since a larger contribution of the fractional energy density in gravitational waves as well as a smaller value of $\mathcal{F}$ would both imply a stronger bound on $\eta$, only the region above this line is compatible with the observed PTA signal. There is then only a tight range of about one order of magnitude in $\eta$ that is compatible with PTA experiments and avoids domain wall domination. Notice that, if one considers networks with $\mathcal{F}<1$ --- as one would realistically expect, since domain walls may also emit scalar radiation in their decay --- this allowed range of $\eta$ becomes even narrower. This means, in particular, that values of $\mathcal{F}\lsim 0.004$ can be ruled out if domain walls are to be the main contributors to the PTA signal.

Additionally, under the assumption that the peak of the domain wall gravitational wave spectrum lies within the frequency interval [$f_{\rm min}$, $f_{\rm max}$], one may see that the allowed range of $z_\star$ is also very narrow for each value of $\alpha$. Fig. \ref{fig:constraintsPTA} shows the region of the ($z_\star$, $\eta$) parameter space compatible with PTA data for $\alpha=1$ and $\alpha=10^{-6}$ (as the opaque dark and light green areas respectively). For an $\alpha$ of order unity, this would imply a decay time of the domain wall network shortly after the QCD phase transitions, while the redshift of decay (roughly) linearly shifts with smaller $\alpha$ to later times as indicated by the color bar on the right side of the plot. Considering the constraints on the time of decay and on  the energy scale $\eta$ at domain wall formation, for a given value of $\alpha$, the region of parameter space consistent with the signal detected by PTAs becomes extremely narrow, requiring a substantial fine tuning of the properties of the domain wall networks in order to explain the SGWB detected by the PTAs.

\section{Discussion and Conclusions}
\label{sec:Conclusion}
In this article we have investigated what are the allowed domain wall scenarios in light of current CMB and PTA data. Since there are some uncertainties regarding the spectrum (especially about the shape of its peak), here we opted to perform our analysis using the total SGWB, instead of its power spectrum, as this may be characterized with minimal assumptions. The main uncertainty in this case is related to the efficiency of GW emission in domain wall collapse, quantified here using the parameter $\mathcal{F}$, but our main conclusions are mostly independent on this parameter.

Our results show that indirect CMB bounds on $\Omega_{\rm gw}$ at the time of decoupling have a limited constraining power on domain wall scenarios and generally provide constraints that are less stringent than those that result from imposing that domain walls provide only a subdominant contribution to the cosmic energy budget. Only if the emission of gravitational radiation, and not scalar radiation, is the dominant decay mechanism of collapsing domain walls --- more precisely if $\mathcal{F}\sim \mathcal{O}(1)$ --- does this CMB bound lead to a decrease of the allowed region of the $(z_\star,\eta$) parameter space (and, even in this case, this decrease is not significant). Notice however that the CMB constraints derived here are conservative because the computation of $\Omega_{\rm gw}$ in Eqs.~(\ref{OGWwalls}) and~(\ref{eq:constraintMatter}) only includes the GWs emitted before the whole domain wall networks decays at $z_\star$. There may also be a significant emission of GWs during the final collapse of the network, which should enhance $\Omega_{\rm gw}$ by up to one order of magnitude~\footnote{In~\cite{Ferreira:2024eru}, the authors find that this decay may be slightly delayed, which may further enhance the amplitude of this background.}. The inclusion of this contribution may then result in a further reduction of the allowed range of parameter space, but this --- unlike the constraints presented here -- is model dependent.

Despite this, we found that even without considering CMB bounds, the range of allowed energy scales for the domain-wall forming phase transition is already quite limited, especially for networks that decay early in cosmic history. For instance, for walls decaying at $z_\star\sim 10^{18}$ (at a temperature of roughly $T_\star\sim 100\,\rm{TeV}$), we should have that $10^2 \lesssim \eta \lesssim 10^6\, \rm{TeV}$ (roughly 4 order of magnitude in range), while for $z_\star \sim 10^8$ ($T_\star\sim 10^{-1}\,\rm{MeV}$), the allowed parameter range is much wider: $10^{-1} \lesssim \eta \lesssim 10^6 \,\rm{MeV}$.

We have also found that, if domain walls are to provide an explanation to the observed PTA signal, this would require a significant fine tuning of the parameters of the model. Domain walls would not only need to have a significant GW emission efficiency --- we found that $\mathcal{F}\gtrsim 10^{-3}$, but, as previously explained, we were quite conservative in deriving this constraint --- but their energy density would have to be very close to dominating the energy density of the universe. As a matter of fact, the allowed range of parameters is very narrow: one should have, roughly, that $\eta \sim \mathcal{O}(10^2)\,\rm{TeV}$ and, for $\alpha=1$, $T_\star \sim \mathcal{O}(10^2)\,\rm{MeV}$~\footnote{$T_\star$ would roughly decrease proportionally to $\alpha$.}. Like in the previous case, this result was derived based on the contribution to the SGWB before the actual decay of the network. Here, in fact taking an enhancement of the SGWB due to the network's decay into account could potentially increase the range of allowed parameter space by lowering the minimal domain wall tension required to reproduce the signal. As discussed before, this would be model dependent and at most lower the minimal $\sigma$ by one order of magnitude, which would translate into less than a factor of $1/2$ in the characteristic energy scale. These results are in very good agreement with those presented in the literature~\cite{Ferreira:2024eru,NANOGrav:2023hvm,Kitajima:2023cek,Zhang:2023nrs,Gouttenoire:2023ftk}, but unlike those, ours are based on minimal assumptions and as model-independent as possible. The results in the literature all stem from numerical simulations of particular domain wall models~\cite{hiramatsuEstimationGravitationalWave2014}, in which the energy density of GWs, $\rho_{\rm gw}$, was measured. These simulations seem to indicate that $\rho_{\rm gw}$  remains constant and is roughly given by $\sim G\sigma^2$ (aside from an efficiency parameter of order unity), which would correspond to having $\mathcal{F}$ growing linearly with time in this framework. Although such a regime may indeed exist (as discussed in detail in~\cite{Gruber:2024dgw}), it would necessarily have to be transient as $\mathcal{F}$ should necessarily be smaller than unity at all times --- after all, the energy emitted in the form of GWs cannot ever exceed the energy lost by the network. As a matter of fact, in such a regime, as $\mathcal{F}$ increases and $\rho_{\rm gw}$ becomes comparable to the domain wall energy density, gravitational backreaction should start having a significant impact on GW emission. Such a regime cannot be captured by numerical simulations and the last stages of collapse --- in which walls becomes ultra-relativistic and the emission GWs may be quite significant~\cite{Gruber:2024dgw} --- cannot as well. This casts some doubt as to the ability of numerical simulations to describe realistic domain wall scenarios. Assuming, as in simulation-inferred models, that $\mathcal{F}\propto t$ results in an additional factor of $G\sigma/H_\star$ in $\Omega_{\rm gw}$ (note that in our model $\rho_{\rm gw}$ scales as $\sigma H_\star$), but despite this our interpretations of PTA data roughly coincide. This coincidence may be easily explained: since, in both cases, domain wall density would have to be comparable to that of the background density at the time of decay, we should have, in both cases, that $\sigma H_\star \sim H_\star^2/G$, which implies that $G\sigma/H_\star\sim 1$. This is precisely the limit in which both approaches agree --- they would not if domain walls were to be subdominant --- but it is also the limit in which the potential impact of gravitational backreaction is most severe.

%%%%%%%%%%%%%%%%%%%%%%%%%%%%%%%%%%%%%%%%%%%%%%%%%%%%%$
\begin{acknowledgments}
The authors would like to thank Dr. Paul Tol for developing the colorblind-friendly color schemes~\cite{PTColorSchemes} used in this work. D. G. is supported by FCT -- Funda\c{c}\~{a}o para a Ci\^{e}ncia e a Tecnologia through the PhD fellowship 2020.07632.BD. L. S. is supported by FCT through contract No. DL 57/2016/CP1364/CT0001. Funding for this work has also been provided by FCT through the research grants UIDB/04434/2020 and UIDP/04434/2020 and through the R \& D project 2022.03495.PTDC -- \textit{Uncovering the nature of cosmic strings}.

\end{acknowledgments}
%%%%%%%%%%%%%%%%%%%%%%%%%%%%%%%%%%%%%%%%%%%%%%%%%%%%%%%%%%

\bibliography{walls}
 	
 \end{document}